\documentclass[12pt,twoside]{article}
\usepackage{fleqn,espcrc1}

\usepackage{graphicx}
\newcommand{\AmS}{{\protect\the\textfont2
  A\kern-.1667em\lower.5ex\hbox{M}\kern-.125emS}}

\title{Nucleosynthesis in massive stars revisited}

\author{T. Rauscher\address{
        Department of Physics and Astronomy,
        University of Basel, Basel, Switzerland}$^{\rm b}$%
        \thanks{Supported by a PROFIL professorship of the Swiss SNF.},
       A. Heger\address{
        Department of Astronomy and Astrophysics,
        UC Santa Cruz, Santa Cruz, CA, USA},
        R. D. Hoffman\address{Lawrence Livermore National Laboratory, 
Livermore, CA, USA},
        and
        S. E. Woosley$^{\rm b}$}

\begin{document}

\maketitle

\begin{abstract}
\noindent
We have performed the first calculations to follow the evolution of all
stable nuclei and their radioactive progenitors in a
finely-zoned stellar model computed from the onset of central hydrogen
burning through explosion as a Type II supernova. Calculations
were done for 15 $M_{\odot}$, 20 $M_{\odot}$, and 25 $M_{\odot}$
Pop I stars using the most recently
available set of experimental and theoretical nuclear data, revised
opacity tables, and taking into account mass loss due to stellar winds.
Here results are presented for one 15 $M_{\odot}$ model.
\end{abstract}

\section{INTRODUCTION}

\begin{figure}[t]
\centerline{\includegraphics*[bb=50 50 792 550,height=18.83pc]{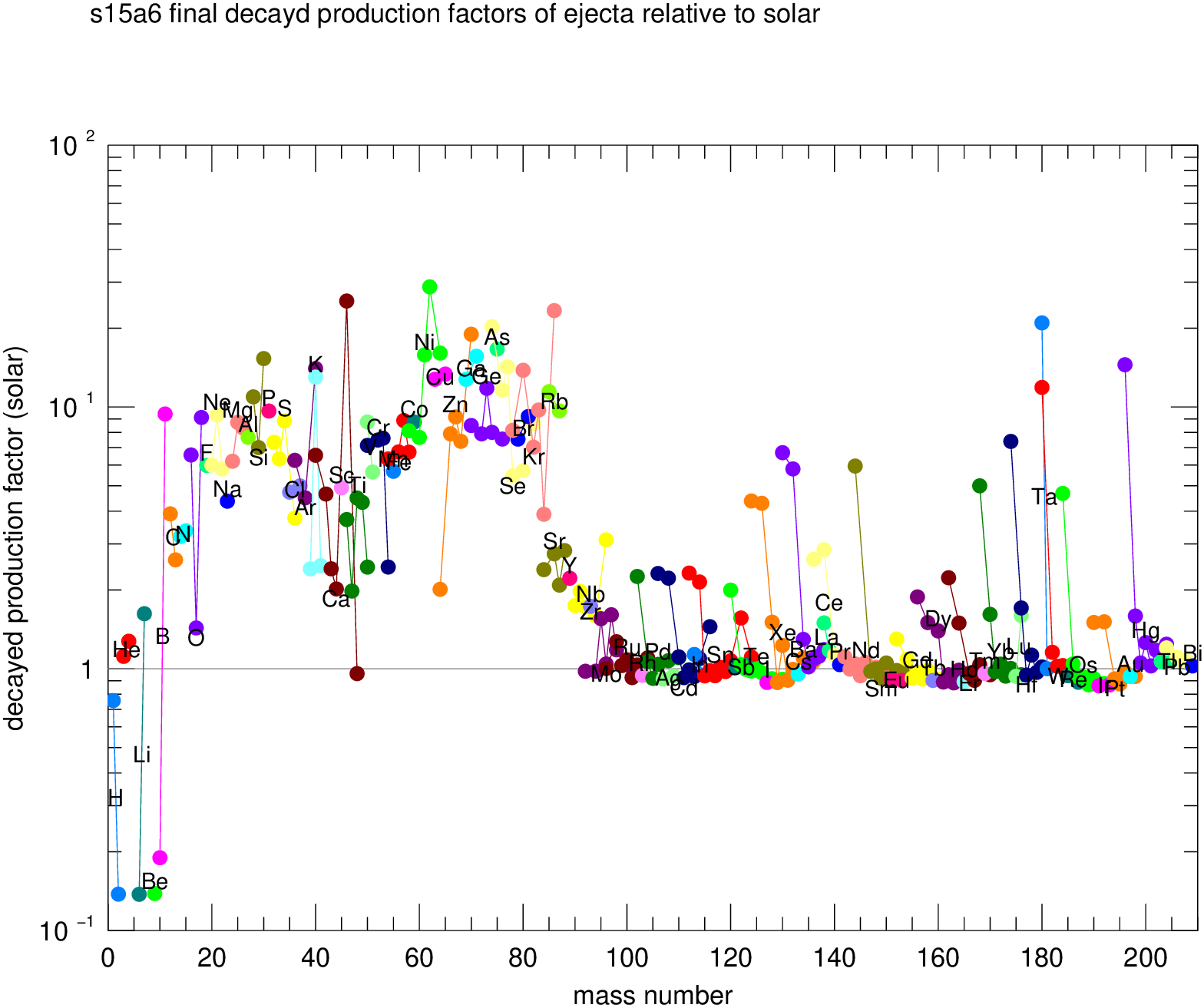}}
\caption{\label{rau:fig} Production factors in the ejecta of a 15
$M_{\odot}$ star relative to solar abundance vs.\ nuclear mass number.}
\end{figure}
Stars above $\sim10$ $M_{\odot}$ are responsible for producing most of the
oxygen and heavier elements found in nature.  Numerous studies have been
devoted to the evolution of such
stars and their nucleosynthetic yields, e.g.,\ \cite{rau:WW95,rau:TNH96}.  
However, our knowledge of both the input
data and the physical processes affecting the evolution of these stars
has improved dramatically in recent years.  Thus, it became
worthwhile to improve on and considerably extend the previous
investigations of pre-- and post--collapse evolution and
nucleosynthesis. We
present the first calculation to determine, self-consistently,
the complete synthesis of all stable nuclides in any model for a
massive star.
Due to the limited space, in this report we mainly focus on giving an
outline of our investigations. For further details on the calculations
and results the reader is referred to the full papers 
\cite{rau:rhhw00,rau:heg00}.

\section{REACTION NETWORK}

We employed a nuclear reaction network of unprecedented size in full
stellar evolution calculations. The
network used by \cite{rau:WW95} (WW95), large in its
day, was limited to 200 nuc\-lides and extended only up to germanium.  
Studies using
reaction networks of over 5000 nuclides have been carried out
for single zones or regions of stars, especially to obtain the
r-process, e.g., \cite{rau:CCT85,rau:fre99,rau:kra93}, but
``kilo-nuclide'' studies of nucleosynthesis in complete stellar models
(typically of 1000 zones each) have been hitherto lacking.
Similar to WW95, nucleosynthesis was followed by co-processing the stellar
model throughout its evolution using the extended nuclear reaction
network.  From hydrogen ignition through central helium depletion a
617 nuclide network was employed that included all elements up to
polonium, adequate to follow the s-process.  Just prior to central carbon
ignition, we switched to a network containing 1482 nuclides.
That network incorporated more neutron-rich isotopes
to follow the high neutron fluxes in carbon (shell) burning and was also
slightly extended on the proton-rich side
to follow the $\gamma$-process
\cite{rau:WH78,rau:ray95}. The nucleosynthesis during
the supernova explosion itself was followed in each zone using a 2437
nuclide network including additional proton-rich isotopes to
better follow the $\gamma$-process in the neon-oxygen core, and also
many additional neutron-rich isotopes to follow the n-process expected
during supernova shock front passage through the helium shell.
Here we will ignore
the nucleosynthesis that occurs in the neutrino wind which may be the
principal site of the r-process because its thermodynamic properties continue
to be poorly understood.

\section{INPUT PHYSICS}

Our calculations were performed using the
stellar evolution code KEPLER \cite{rau:WW95} with several
modifications relative to WW95 
(mass loss due to stellar winds \cite{rau:NJ90}, improved
adaptive network) and updates (OPAL95 opacity tables
\cite{rau:IR96}, neutrino loss rates \cite{rau:ito96}).
We generated a new library of experimental and theoretical reaction
rates. As the basis of our reaction rate set we used
statistical model calculations obtained with the NON-SMOKER
code \cite{rau:rtk97,rau:RT00}. A library of theoretical reaction rates
calculated with this code and fitted to an analytical function --- ready to be
incorporated into stellar model codes --- was published recently
\cite{rau:RT00}. It includes
rates for all possible targets from proton to neutron dripline and between
Ne and Bi, thus being the most extensive published library of theoretical
reaction rates to date. For the network described here we
utilized the rates based on the FRDM set.
This was supplemented with experimental neutron capture rates
along the line of stability \cite{rau:bao00}.
Experimental ($\alpha$,$\gamma$) rates were
implemented for $^{70}$Ge
\cite{rau:fue96} and $^{144}$Sm \cite{rau:som98}. The derived
$\alpha$+$^{70}$Ge and $\alpha$+$^{144}$Sm potentials were also
utilized to recalculate the transfer reactions involving these potentials.
For the important rates for $\alpha$-capture reactions on self-conjugated
($N=Z$) nuclides, a new semi-empirical rate determination 
was implemented \cite{rau:rtgw00}.
For comparison, we used different sets of experimental
and theoretical rates for elements below neon: WW95, Ref.\ \cite{rau:HWW00},
and NACRE \cite{rau:NACRE}. 
%For the important rate 
%$^{12}$C($\alpha$,$\gamma$)$^{16}$O we used an updated rate
%($S(300)=146$ keV barn) and temperature dependence
%\cite{rau:buc00}.
Experimental $\beta^-$, $\beta^+$, and $\alpha$-decay rates were taken from
\cite{rau:NWC95}
and theoretical $\beta^-$ and $\beta^+$ rates from
\cite{rau:moe96}.  As a special case, we implemented a
temperature-dependent $^{180}$Ta decay
\cite{rau:end99}. For $A\leq 40$ we also included recent
theoretical weak rates \cite{rau:LM00}.
We did not follow the $\nu$-process for nuclides with $Z$ or $N$
larger than $40$.
The supernova explosion was simulated, as in \cite{rau:WW95}, by an
inward--outward moving piston 
resulting in a total kinetic energy of the ejecta at infinity of
1.2$\times 10^{51}$ erg. The final mass cut outside
the piston is determined self-consistently from the hydrodynamic
calculation.

\section{RESULTS AND DISCUSSION}

Here we only summarize the important
results for one 15 $M_{\odot}$ Pop I star. The production factors of this
non-rotating model are shown in
Fig.\ \ref{rau:fig} as an example of our results.
The model included mass loss and the rate set of \cite{rau:HWW00} below
Ne.
Note, however, that though this
mass is a numerically typical case of a Type II or Ib supernova, the
average nucleosynthetic yield of massive stars is the result of
populations of different stars each of which has its own peculiar
yields which must be combined to result in a solar-like abundance
pattern. Other stars and rate sets will be discussed in \cite{rau:rhhw00}.

\subsection{Stellar structure}

The revision of the opacity table and the introduction of mass loss
generally leads to smaller helium core sizes which tend to also
decrease the mass of the carbon-oxygen and the silicon core.  
Note, however, that the absolute values of
these core masses depend on the uncertainties, in particular, of the
mixing processes in the stellar interior, such as semiconvection,
overshooting, and rotation.

The change in the weak rates \cite{rau:LM00}, important after
central oxygen burning, leads to a $2-3\,\%$ higher electron fraction
per nucleon, $Y_{\rm e}$, at the time of core collapse in the center of the
star and the ``deleptonized core'' tends to
comprise less mass \cite{rau:heg00}.  More important for the core
collapse supernova mechanism might be the $30-50\,\%$ higher densities
of the new models between the region of $m=1.5-2$ $M_{\odot}$
\cite{rau:heg00}, which may result in a correspondingly higher
ram-pressure of the infalling matter.

\subsection{Intermediate and heavy element nucleosynthesis}

A strong secondary s-process contribution appears between iron and a
mass number of $A=90$.  Above $A=100$ the s-process in our 15 $M_{\odot}$
star is very weak, but it becomes notably stronger in stars with more
massive helium cores that perform helium burning at higher entropies.
Furthermore, the strength of the s-process is found to be very sensitive to the 
($\alpha$,n)--($\alpha$,$\gamma$) rates and branching on $^{22}$Ne which is
experimentally not well determined (see also \cite{rau:ray00}).
Second only to the well-known strong dependence of the stellar structure on the
$^{12}$C($\alpha$,$\gamma$) rate, it becomes another important candidate
for further laboratory study.

The proton-rich heavy isotopes above $A=123$ can be well produced by
the $\gamma$-process occurring during implosive and explosive oxygen
and neon burning. The proton-rich isotopes around $A=160$ and those
between $A=100$ and $A=123$, however, are underproduced by a factor of
$3$ to $4$ with respect to $^{16}$O. The isotope $^{180}$Ta seems to
show a strong
overproduction by the $\gamma$-process. However, in the figure only the
totally produced $^{180}$Ta is shown. 
The surviving yield can
be brought down to a more consistent production level
by accounting for the
distribution between ground and isomeric states \cite{rau:rhhw00}.

The expected r- or n-process production by the supernova shock
front passing through the base of the helium shell is not
significant in any of our model stars, not
even at $A=130$. We observed some redistribution of isotopes at the
base of the helium shell around $A=123$ but this did not show the
characteristics of a typical r-process nor was it important compared
to the total yield of the star.

Summarizing, we have presented the first calculation to follow the complete
s-process through all phases of stellar evolution and the
$\gamma$-process in the whole star through the presupernova stage and
subsequent supernova explosion.
This research was supported, in part, by DOE
(W-7405-ENG-48), NSF (AST 97-31569, INT-9726315), the Alexander von
Humboldt Foundation (FLF-1065004), and the Swiss NSF (2124-055832.98).

%----------------------------------------------------------------------

\end{document}